\documentclass[a4paper,11pt]{article}
\pdfoutput=1 

\usepackage{jinstpub} 
\usepackage{tablefootnote}
\usepackage{lineno}

\title{\boldmath KM3NeT sensitivity to low energy astrophysical neutrinos}

\author[a]{Gwenha\"el de Wasseige\note{Corresponding author.}}

\affiliation[a]{Laboratoire APC, Universit\'e de Paris\\10, Rue Alice Domon et L\'eonie Duquet, 75013 Paris, France}

\emailAdd{gdewasseige@km3net.de}

\abstract{KM3NeT, a new generation of neutrino telescope, is currently being deployed in the Mediterranean Sea. While its two sites, ORCA and ARCA, were respectively designed for the determination of neutrino mass hierarchy and high-energy neutrino astronomy, this contribution presents a study of the detection potential of KM3NeT in the MeV-GeV energy range. At these low energies, the data rate is dominated by low-energy atmospheric muons and environmental noise due to bioluminescence and K-40 decay. The goal of this study is to characterize the environmental noise in order to optimize the selection of low-energy neutrino interactions and increase the sensitivity of KM3NeT to transient astrophysical phenomena, such as close-by core-collapse Supernovae, solar flares, and extragalactic transients. In this contribution, we will study how using data science tools might improve the sensitivity of KM3NeT in these low-energy neutrino searches. We will first introduce the data sets and the different variables used to characterize KM3NeT's response to the environmental noise. We will then compare the efficiency of various tools in identifying different components in the environmental noise and in disentangling low-energy neutrino interactions from the background events. We will conclude with the implication of low-energy neutrinos for future astrophysical transient searches.}

\collaboration[c]{on behalf of the KM3NeT Collaboration}

\proceeding{Very Large Volume Neutrino Telescope 2021\\
18-21 May 2021\\
Valencia, Spain}

\begin{document}
\maketitle
\flushbottom

\section{Motivation}
\label{sec:intro}

The KM3NeT neutrino telescope is currently being deployed in the Mediterranean Sea. The two different sites, ORCA and ARCA located off-shore Toulon and Portopalo di Capo Passero, respectively, are optimized to detect neutrinos in different energy ranges: while the large instrumented volume of ARCA makes it an ideal tool to detect neutrinos with TeV energies and above, ORCA with its dense instrumentation has been thought to detect neutrinos from 3~GeV and above. Although the primarily goal of ORCA is to determine neutrino mass hierarchy and measure neutrino oscillation parameters, its layout also allows us to search for low-energy astrophysical neutrinos at the sub-GeV level. KM3NeT is hence sensitive to astrophysical neutrinos across a very wide energy range using a single detection unit technology, a Digital Optical Module (DOM) hosting 31 3-inch photomultiplier tubes (PMTs)~\cite{loi}.  There currently exists a gap in the energy coverage of KM3NeT between the 100~MeV range, where the supernova analysis~\cite{sn_paper} stops being efficient, and the few GeV range where the standard ORCA neutrino search starts to become valid~\cite{online}. Less than 15\% of neutrino interactions falling in this energy gap  pass the current ORCA trigger configuration. The goal of the work presented in these proceedings is to carry out preliminary investigations to determine whether KM3NeT could gain sensitivity in this energy gap. While a search for a steady astrophysical neutrino signal is challenging due to the continuous environmental background recorded by the DOMs, a search for neutrinos from transient and variable sources may be feasible if the background can be sufficiently reduced. The dominant background in this energy range is expected to come from the environment, more specifically from bioluminescence and the decay of Potassium-40 (K-40). In the following, we present an approach allowing us to characterize the signature of the environment to then disentangle it from GeV neutrinos. We show how the approach is affected by different environmental conditions and detector sites. We conclude by showing preliminary promising results and sketching the path forward.

\section{Description of the adopted approach}

In order to maximize the detection volume, and hence the sensitivity to a transient sub-GeV flux, we have chosen to focus on the signature recorded on single DOMs. This will allow us to apply similar event selection criteria to both ORCA and ARCA DOMs, significantly increasing the effective area available for detection. We search for specific patterns in time (t) and space (x,y,z) recorded by the PMTs on each DOM. We use minimum bias data recorded periodically by the ORCA and ARCA lines already taking data, to remove any impact of the current trigger configuration on the studied events. Using GENIE-based simulations of sub-GeV neutrino interactions~\cite{gseagen}, we estimate the time window in which all the hits due to the interactions are recorded by the PMTs to be of 30~ns. For each time slices of 30~ns in the mininum bias data, as well as in GeV neutrino and atmospheric muon simulated events,  we derive several variables, presented in Table~\ref{table:variables}. It has been verified that these variables are correctly simulated by comparing their overall distributions and quartiles in the minimum bias data with those of atmospheric muon and noise simulations, which are the dominant components in data when no selection is applied. Furthermore, it has been verified that none of these variables show a strong correlation with each other.

\begin{table}
\centering
\caption{List of variables considered in the analysis.}
\begin{tabular}{|c|c|}
\hline
Variables integrated over 30~ns on the DOM& Variables in between consecutive hits on the DOM \\
\hline
\hline
Number of hits & Mean angle between PMTs\\
Sum of the Time over Threshold (ToT)~\tablefootnote{The time over threshold is a proxy of the charge recorded by the PMT.} & Standard deviation of the angle between PMTs \\
Mean PMT ID~\tablefootnote{Each PMT is assigned a number, from the bottom to the top. The mean PMT ID is therefore gives an estimate of the position on the DOM where most hits have been recorded.} & Mean PMT ID\\ 
Number of hits on the bottom part of the DOM& Standard deviation of the PMT ID distribution\\
Number of hits on the top part of the DOM& Mean time difference\\
Mean time difference between the hits and the first hit& \\
Standard deviation of the time difference distribution & \\
\hline
\end{tabular}
\label{table:variables}
\end{table}
This exploratory analysis uses data science tools to build new variables based on the original variables presented in Table~\ref{table:variables}.  Three different tools have been used: Principal component analysis (PCA), Isomap, and Stochastic neighbour embedding method (t-SNE). Each of these tools combine the original variables to build two new variables, named hereafter $(z_1,z_2)$, while minimizing a cost function to guarantee the new phase space to be as representative of the original one as possible. The new variables are linear or non-linear combinations of the original ones.  When using the newly built variables, we conserve a good data-simulation agreement.

\section{Results of preliminary investigations}
\begin{figure}[htbp]
\centering 
\includegraphics[width=.42\textwidth]{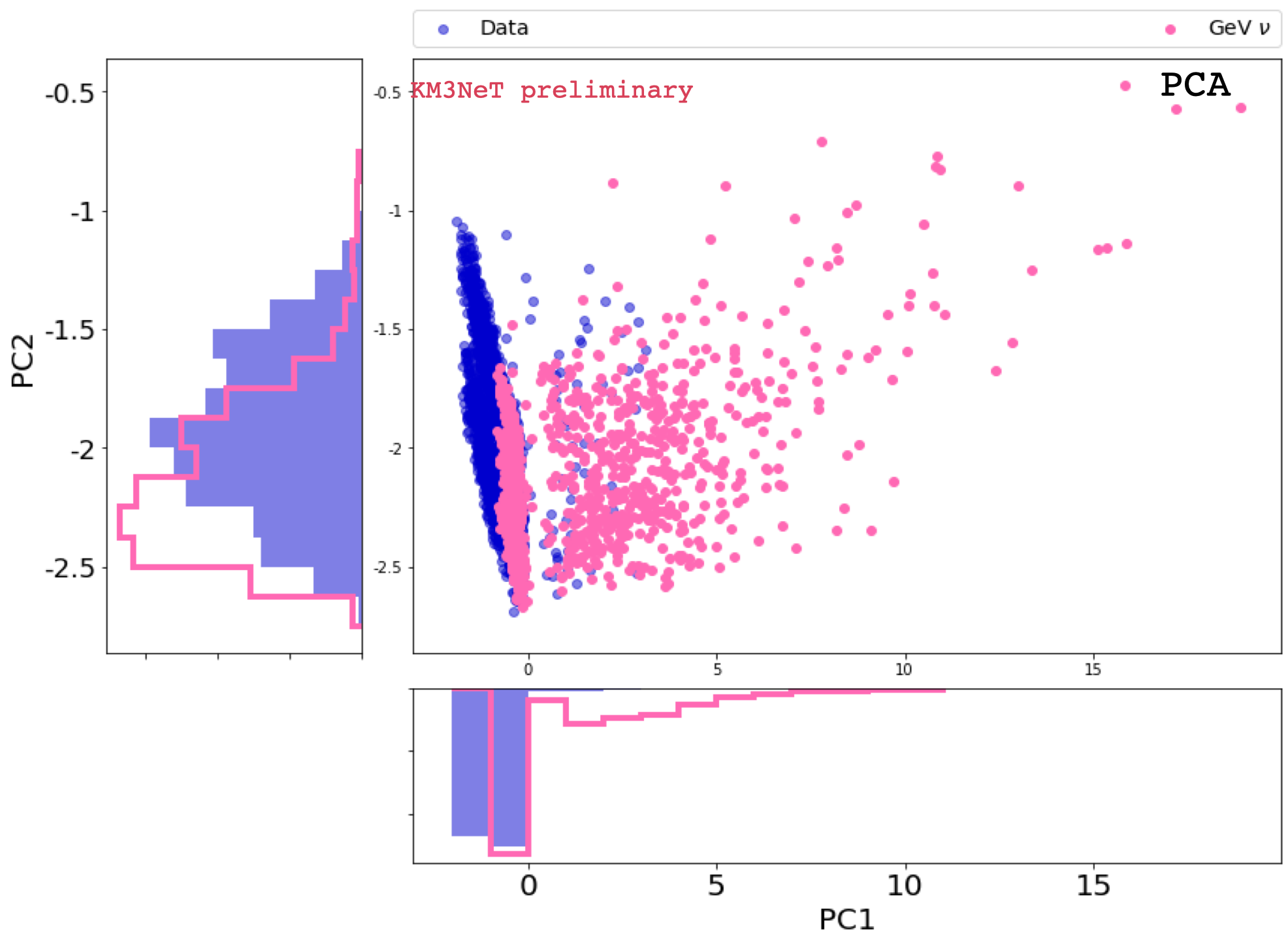}
\qquad
\includegraphics[width=.4\textwidth]{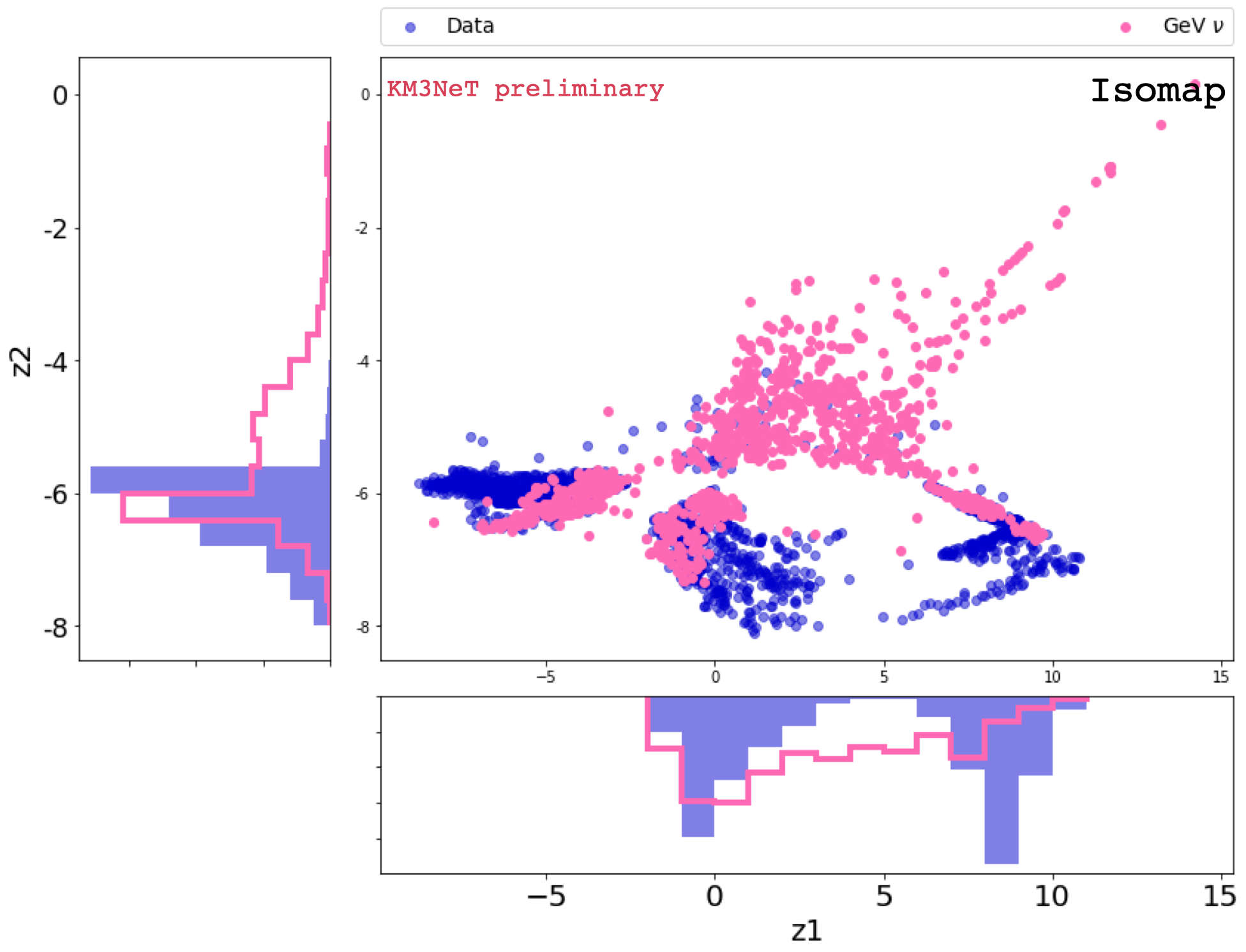}
\qquad
\includegraphics[width=.4\textwidth]{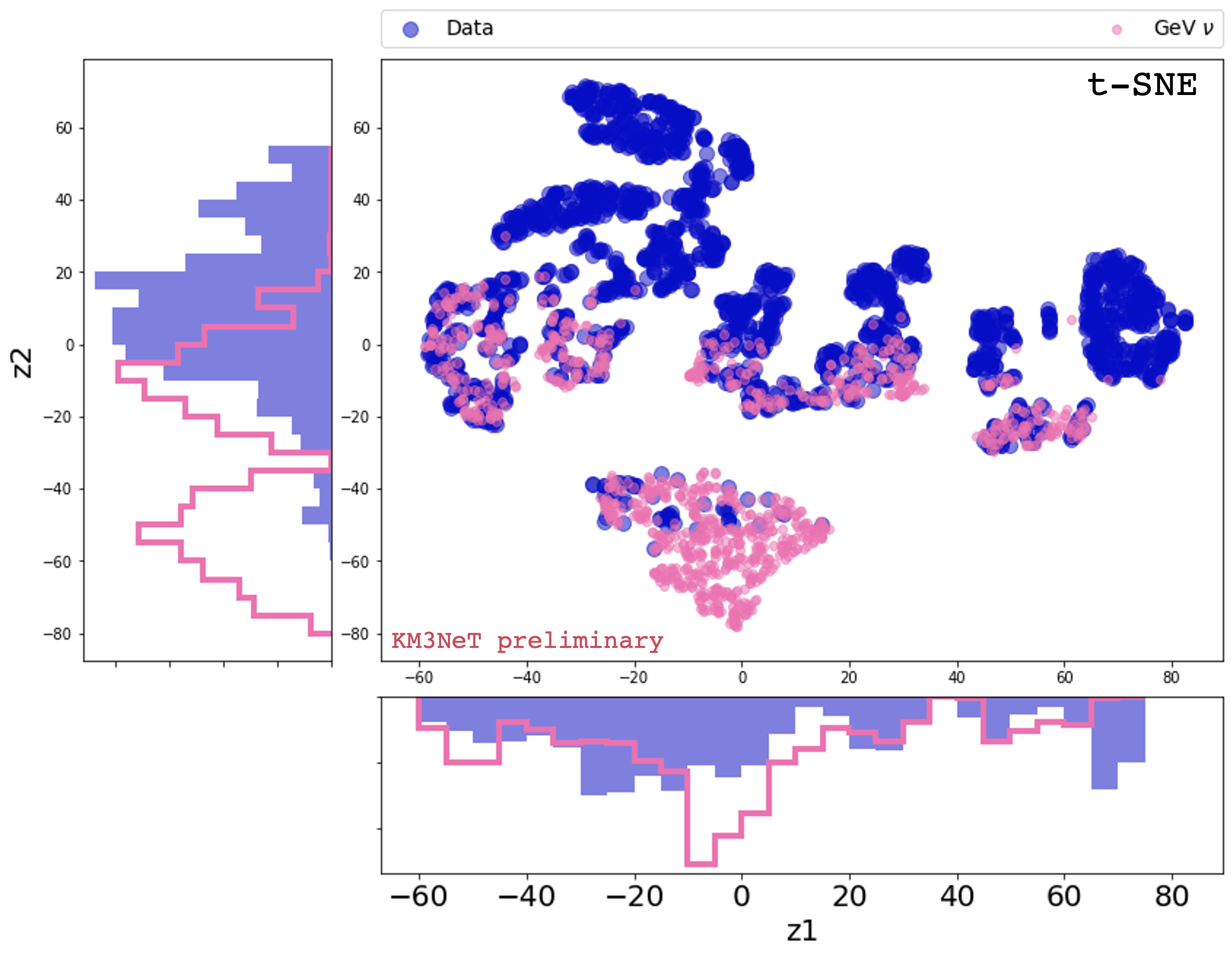}
\caption{\label{fig:data_nu} Distribution of data (blue) and sub-GeV neutrino (pink) events in the PCA (left), Isomap (right), and t-SNE (bottom) $(z_1,z_2)$ space.}
\end{figure}

In order to evaluate the impact of different data taking conditions on the potential selection criteria we could develop, we search for variations in the data distribution in the  $(z_1,z_2)$ space. No significant difference is found when comparing ORCA and ARCA data nor when comparing different level of bioluminescence activity in the detector. 
We then run the three data science tools on minimum bias data and sub-GeV neutrino simulated events. The obtained results are shown in Fig.~\ref{fig:data_nu}, where data is shown in blue and GeV neutrinos are represented in pink in the $(z_1,z_2)$ space. While only minor differences between the two distributions can be found in the case of PCA, both Isomap and t-SNE create some islands in the data distribution where a fraction of the GeV neutrinos is isolated from the data. 
For illustration purpose, a naive cut $(z_2 < -30)$, selecting the southern island in the t-SNE distribution shown in the bottom panel of Fig.~\ref{fig:data_nu}, allows us to keep about 50\% of the GeV neutrino interactions and only 2\% of the minimum bias data. It appears therefore feasible to develop new selection trigger conditions in ORCA and ARCA to increase the fraction of sub-GeV neutrino interactions that are saved and therefore fill the current gap in the KM3NeT energy coverage.


\section{Summary and outlook}
We carry out an exploratory analysis attempting to bridge the energy gap between 100~MeV and a few GeV currently present in the coverage of ORCA and ARCA. This energy range is currently dominated by environmental noise and a precise understanding of the signature left by bioluminescence and K-40 decay is needed to disentangle it from sub-GeV neutrino interactions. After building a list of variables sensitive to different signatures in time and space recorded by the PMTs on a single DOM, we use data science tools to create a new variable space of lower dimension. A very good data/simulation agreement is found both when using the original and new variables. When tested for different data taking conditions, the new variables do lead to significant differences in the obtained data distribution. 
We then compare the distribution of data and sub-GeV neutrino interaction in the new variable space. It has been shown that t-SNE was particularly interesting to disentangle GeV neutrinos from background. 
We will now work on the optimization of the cuts, including removing high energy events coming from atmospheric muons and neutrinos, and derive the sensitivity of KM3NeT to astrophysical transients in the sub-GeV range.

\acknowledgments
G. de Wasseige acknowledges support from the European Union's Horizon 2020 research and innovation programme under the Marie Sklodowska-Curie grant agreement No 844138.

\end{document}